\documentclass[useAMS,usenatbib]{mn2e}

\usepackage{graphicx}
\usepackage{CJKutf8}
\usepackage{amssymb,amsmath}
\newif\ifAMStwofonts
\AMStwofontstrue

\newcommand{\be}{\begin{equation}}
\newcommand{\ee}{\end{equation}}
\newcommand{\bea}{\begin{eqnarray}}
\newcommand{\eea}{\end{eqnarray}}

\def\ltsima{$\; \buildrel < \over \sim \;$}
\def\simlt{\lower.5ex\hbox{\ltsima}}
\def\gtsima{$\; \buildrel > \over \sim \;$}
\def\simgt{\lower.5ex\hbox{\gtsima}}

  \newcommand{\bc}{\begin{center}}
  \newcommand{\ec}{\end{center}}
  \newcommand{\Msun}{~{\rm M_\odot}}
  \newcommand{\hMsun}{~h^{-1}\>{\rm M_\odot}}
  \newcommand{\Mpc}{~h^{-1}~{\rm Mpc}}

  \newcommand{\vel}{\,{\rm km\,s^{-1}}}

\defcitealias{Cui2012a}{Paper I}	

\title[The effect of AGN feedback on the HMF]{The effect of AGN feedback on the
  halo mass function} 

\author[W.~Cui, S. Borgani, G. Murante]{Weiguang Cui$^{1,2,3}$ 
  \thanks{E-mail:weiguang.cui@uwa.edu.au}, Stefano Borgani$^{3,4,5}$ \&
  Giuseppe Murante$^3$\\
$^1$ ICRAR, University of Western Australia, 35 Stirling Highway,
Crawley, Western Australia 6009, Australia\\
$^2$ ARC Centre of Excellence for All-Sky Astrophysics (CAASTRO)\\
$^3$ Astronomy Unit, Department of Physics, University of Trieste,
via Tiepolo 11, I-34131 Trieste, Italy\\
$^4$ INAF -- Astronomical Observatory of Trieste, via Tiepolo 11,
I-34131 Trieste, Italy\\
$^5$ INFN -- Sezione di Trieste, I-34100 Trieste, Italy}

\begin{document}

%\date{Accepted ???. Received ???; in original form }

\pagerange{\pageref{firstpage}--\pageref{lastpage}} \pubyear{2014}

\maketitle

\label{firstpage}
\begin{CJK*}{UTF8}{gkai}
\begin{abstract}
  We investigate baryon effects on the halo mass function (HMF), with
  emphasis on the role played by AGN feedback. Halos are identified with
  both Friends-of-Friends (FoF) and Spherical Overdensity (SO)
  algorithms. We embed the standard SO algorithm into a
  memory-controlled frame program and present the {\bf P}ython
  spher{\bf I}c{\bf A}l {\bf O}verdensity code --- {\small PIAO}
  (Chinese character: 漂). 

  For both FoF and SO halos, the effect of AGN feedback is that of
  suppressing the HMFs to a level even below that of Dark Matter
  simulations. The ratio between the HMFs in the AGN and in the DM
  simulations is $\sim 0.8$ at overdensity $\Delta_c=500$, a
  difference that increases at higher overdensity $\Delta_c=2500$,
  with no significant redshift and mass dependence. 
  A decrease of the halo masses ratio with respect to the DM case induces
  the decrease of the HMF in the AGN simulation. The shallower inner density
  profiles of halos in the AGN simulation witnesses that mass
  reduction is induced by the sudden displacement of gas
  induced by thermal AGN feedback. We provide fitting functions to
  describe halo mass variations at different overdensities, which
  can recover the HMFs with a residual random scatter $\simlt 5$ per
  cent for halo masses larger than $10^{13} \hMsun$.
\end{abstract}
\end{CJK*}

%*****************************************************************************

\section{Introduction}
\label{i}

The halo mass function (HMF hereafter) is a unique prediction of
cosmological models of structure formation. 
The evolution of the HMF traced by galaxy clusters has been recognized
since a long time as a powerful tool to trace the growth of cosmic
structures and, therefore, to constrain cosmological parameters
\cite[see][for reviews, and references
therein]{Rosati2002,Allen2011}. In particular, cosmological
applications of the HMF require to know its shape and evolution to a
high precision, in order to fully exploit its potential as a
cosmological tool to be applied to ongoing and future large surveys of
galaxy clusters \citep[e.g.][]{Wu2010,Murray2013}.

N--body simulations covering wide dynamic ranges are nowadays
providing rather accurate calibration of the mass function of Dark
Matter (DM) halos \citep[e.g.][]{Jenkins2001, Reed2003, Reed2007,
  Reed2013,Lukic2007,Tinker2008,Crocce2010,Courtin2011,
  Bhattacharya2011,Angulo2012,Watson2013}. Various extensions of the
standard $\Lambda$CDM cosmology model, such as coupled dark energy
models \citep[e.g.][]{Cui2012b, Baldi2012}, modified gravity models
\citep[e.g.][]{Schmidt2009,Zhang2013,Puchwein2013}, non-Gaussian
initial conditions \citep[e.g.][]{Grossi2009,Pillepich2010}, massive
neutrinos \citep[e.g.][]{Brandbyge2010,Ichiki2012,Costanzi2013}, Warm
Dark Matter \citep[e.g.][]{Schneider2013,Angulo2013}, have been
studied using numerical simulations, and their effect on the HMF has
been investigated.

A crucial aspect in the 
calibration of the HMF is related to the algorithm used to identify
halos, the two most widely used being the Friend of Friend (FoF) one
and the Spherical Overdensity (SO) one. The choice of a specific
algorithm clearly impacts both the number of identified halos and
their mass \citep[e.g.][; see also \citealt{Knebe2011} for a
detailed comparison between different halo finders]{White2001,
  White2002, Lukic2009, More2011, Watson2013}.

All the above mentioned HMF calibrations are based on N--body
simulations that follow the evolution of a collisionless DM fluid. On
the other hand, the presence of baryons is known to add subtle but
sizeable effects on halo formation and internal structure, whose
details also depend on the physical processes included in the
numerical treatment of the baryonic component, such gas
cooling, star formation and energy feedback from supernovae (SN) and
Active Galactic Nuclei (AGN) \citep[e.g.][and references
therein]{KravtsovBorgani2012}.

A number of studies based on cosmological hydrodynamical simulations
have been recently carried out to analyse in detail the effect of
baryonic processes on different properties of the total mass
distribution, such as the power spectrum of matter density
fluctuations \citep[e.g.]{Rudd2008, Daalen2011, Casarini2012}, the
halo correlation functions \citep[e.g.][]{Zhu2012, Daalen2013}, the
halo density profiles \citep[e.g.][]{Duffy2010, Lin2006} and
concentration \citep[e.g.][]{Rasia2013, Bhattacharya2013}, and the HMF
\citep[e.g.][]{Stanek2009, Cui2012a, Sawala2013, Martizzi2013,
Cusworth2013, Balaguera2013, Wu2013}.

As for the effect of non--radiative hydrodynamics, the presence of
baryons has been shown to induce a slight increase of the HMF
\citep[][hereafter Paper I]{Cui2012a}. When extra--heating is included,
\cite{Stanek2009} found instead a decrease in the HMF. As for the
effect of radiative cooling, star formation and SN feedback, different
groups consistently found an increase of the HMF, an effect that is
more evident in the high--mass end
\citep{Stanek2009,Cui2012a,Martizzi2013}. On the other hand,
\citet{Sawala2013} found that efficient SN feedback produces an
opposite effect in low--mass halos.

On the other hand, a number of analyses have shown in the last years
that including AGN feedback in cosmological simulations provides
populations of galaxy clusters in better keeping with observational
results \citep[e.g.][]{Puchwein2008, ShortThomas2010, Fabjan2010,
  McCarthy2011, Planelles2013, LeBrun2014}. 
When the AGN feedback is included, different results were found by
\citet{Martizzi2013} and \citet{Cusworth2013}. The former showed that
the HMF with AGN feedback is higher than the fitting function from
\citet{Tinker2008}, while the latter predicted a lower HMF compared to
the same fitting function. However, their implementation of the AGN
feedback differ.  \cite{Martizzi2013} described AGN feedback by
computing explicitly gas accretion rates onto super-massive black
holes (SMBHs) included as sink particles in simulations that also
include radiative cooling, star formation and SN feedback
\citep[e.g.][]{Springel2005,BoothSchaye2009}.  \cite{Cusworth2013}
included AGN feedback by computing the associated feedback energy from
the semi--analytic model of galaxy formation by \cite{Guo2011},
without including radiative cooling and assuming zero mass for gas
particles, so that no back-reaction of baryons on the DM distribution
is allowed.

In this paper we extend our previous analysis of baryonic effects on
the HMF, presented by \citetalias{Cui2012a}, by also including in our
simulations the effect of AGN feedback. We directly compare the HMF
obtained from DM--only simulations to those produced by radiative
hydrodynamical simulations both with and without AGN feedback, using
exactly the same initial conditions, mass and force resolutions. The
plan of the paper is as follows. In section \ref{simulation}, we
present the simulations analysed in this paper. Section \ref{halo} is
devoted to the description of the halo identification methods. In
section \ref{results} we present the results of our analysis and
describe in detail the differences in the HMF induced by different
feedback models. Our results are discussed and summarised in Section
\ref{concl}.

\section{The Simulations}
\label{simulation}

Three large--volume simulations are analysed in this paper, namely two
hydrodynamical simulations which include different description of
feedback processes affecting the evolution of baryons and one N--body
simulation including only DM particles. Initial conditions for these
simulations are the same as described in \citetalias{Cui2012a} and we refer
to that paper for further details. The hydrodynamical simulations have
the same number dark matter particles ($1024^3$) and gas particles
($1024^3$). A first hydrodynamical simulation includes radiative
cooling, star formation and kinetic SN feedback (CSF hereafter), while
the second one also includes the effect of AGN feedback (AGN
hereafter). As for the DM simulation, it starts for the same initial
conditions as the hydrodynamical simulations, with the gas particles
replaced by collisionless particles, so as to have the same
description of the initial density and velocity fields as in the
hydrodynamical simulations.
 
The three simulations have been carried out using the TreePM-SPH code
{\small GADGET-3}, an improved version of the {\small GADGET-2} code
\citep{Gadget2}. 
Gravitational forces have been computed using a Plummer--equivalent
softening which is fixed to $\epsilon_{Pl}=7.5h^{-1}$ physical kpc
from $z=0$ to $z=2$, and fixed in comoving units at higher
redshift. The simulations assume flat $\Lambda$CDM cosmology with
$\Omega_{\rm m} = 0.24$ for the matter density parameter, $\Omega_{\rm
  b} = 0.0413$ for the baryon contribution, $\sigma_8=0.8$ for the
power spectrum normalisation, $n_{\rm s} = 0.96$ for the primordial
spectral index, and $h =0.73$ for the Hubble parameter in units of
100$\vel {\rm Mpc}^{-1}$. Initial conditions have been generated at
$z=49$ using the Zeldovich Approximation for a periodic cosmological
box with comoving size $L=410\Mpc$.
The masses of gas and DM particles have a ratio such that to reproduce
the cosmic baryon fraction, with $m_g\simeq 7.36 \times 10^{8} \hMsun$
and $m_{DM}\simeq 3.54 \times 10^{9} \hMsun$, respectively.

In the hydrodynamical simulations, radiative cooling is computed for
non--zero metallicity using the cooling tables by
\cite{sutherland_dopita93}, also including heating/cooling from a
spatially uniform and evolving UV background. Gas particles above a
given threshold density are treated as multi-phase, so as to provide a
sub–resolution description of the inter–stellar medium, according to
the model described by \cite{springel_hernquist03}.
Conversion of collisional gas particles into collisionless star
particles proceeds in a stochastic way, with gas particles spawning a
maximum of two generations of star particles. We also include a
description of metal production from chemical enrichment contributed
by SN-II, SN-Ia and AGB stars, as described by
\cite{tornatore_etal07}. Kinetic feedback is implemented by mimicking
galactic ejecta powered by SN explosions, with a wind mass upload
proportional to the local star-formation rate, $\dot M_w=\eta \dot
M_*$. In the CSF simulation we use $\eta=2$ and $v_w = 500\, {\rm
  km}\, s^{-1}$ for the wind velocity, which corresponds to assuming
about unity efficiency for the conversion of energy released by SN-II
into kinetic energy for the adopted Salpeter IMF.

As for the AGN simulation, it includes both the effect of galactic
winds, with $v_w = 350\, {\rm km}\, s^{-1}$ and the same mass--load
parameter $\eta=2$, along with energy feedback
resulting from gas accretion onto SMBHs. The model of AGN feedback
used in this simulation is the same as that adopted by
\cite{Fabjan2010} and is largely inspired to the model originally
introduced by \cite{Springel2005b}. SMBHs, seeded with an initial mass
of $10^6M_\odot$ in halos resolved with at least 100 DM particles,
subsequently grow by merging with other BHs and by gas accretion. The
latter proceeds at the Bondi rate and is Eddington--limited. A
fraction $\epsilon_r=0.1$ of accreted mass is converted into
radiation, with a fraction $\epsilon_f$ of this radiation  thermally
coupled to the surrounding gas. We assume $\epsilon_f=0.1$ which
increases by a factor of four whenever accretion takes place at a rate
of at most one-hundredth of the Eddington limit.  

We note that the main motivation for efficient SN feedback with $v_w =
500\, {\rm km}\, s^{-1}$ in the CSF simulations lies in the need of
reconciling simulation predictions on the cosmic star formation rate
with observations, at least at redshift $z>2$, a choice that still
produces too efficient star formation at lower redshift
\citep[e.g.][]{Tornatore10}. Although AGN feedback is motivated by the
need of reducing the star formation rate at lower redshift, still its
effect is quite significant already at $z\sim2$. Therefore, in order
to prevent too strong a reduction of star formation around this
redshift when SN and AGN feedbacks are both included, we decided to
reduce by a factor of two the kinetic energy associated to the
former. This lowers the resulting wind velocity to $v_w = 350\, {\rm
  km}\, s^{-1}$.

\section{Halo identification}
\label{halo}

The two most common methods used for halo identification in
simulations are the Friend-of-Friend (FoF) algorithm
\citep[e.g.][]{Davis1985} and the spherical overdensity (SO) algorithm
\citep{Lacey1994}. The FoF algorithm has only one parameter, $b$,
which defines the linking length as $b l$ where $l=n^{-1/3}$ is the
mean inter-particle separation, with $n$ the mean particle number
density. In the SO algorithm, there is also only one free parameter,
namely the overdensity $\Delta_c$. The overdensity determines the
aperture of the sphere, within which the total mean density is
$\Delta_c ~ \rho_{crit}$. Here, $\rho_{crit}$ is the critical cosmic
density. Each of the two halo finders has its own advantages and
shortcomings \citep[see more details in][and references
  thereon]{Jenkins2001, White2001, 
  Tinker2008}, and the differences between the two methods in terms of
halo masses and HMFs have been discussed in several analysis
\citep[e.g.][]{White2002, Reed2003, Reed2007, Cohn2008, More2011,
  Anderhalden2011, Knebe2013, Watson2013}. We adopt both methods to
identify halos in this paper.

\subsection{Friend-of-Friend Halos}

In our three simulations FoF halos are identified by a on-the-fly FoF
finder, with a slight smaller linking length $b=0.16$ compared to
commonly used one, $b = 0.2$. Dark matter particles are linked
first. Then, each gas and star particle is linked to the nearest
dark matter particle, whenever the linking criterion is satisfied.

\subsection{Spherical Overdensity Halos -- {\small PIAO}}

\begin{CJK*}{UTF8}{gkai}
We carry out a spherical overdensity (SO) halo search by using an
efficient memory-controlled parallel {\bf P}ython spher{\bf I}c{\bf
  A}l {\bf O}verdensity halo finding code --- {\small PIAO} (Chinese
character: 漂). This code is based on the standard SO algorithm. Its
aim is not to provide a new halo identification method, but to analyse
large simulations on a small computer server or PC with limited
memories. To overcome a memory deficiency problem, we adopt a simple
strategy, which is based on splitting the whole simulation box into
small mesh-boxes, and analysing them one-by-one. The details of this
strategy and how to incorporate it within the SO method is discussed
in Appendix \ref{A:Piao}.  {\small PIAO} is parallelised with a python
MPI package (MPI4py) to speed up the calculation by taking advantage
of multi-core CPUs.
\end{CJK*} 

We applied {\small PIAO} to the three simulations analysed in this
paper. For all of them, SO halos are identified at three different
overdensity values\footnote{In the following, the overdensity value
  $\Delta_c$ is expressed in units of the cosmic critical density at a
  given redshift, $\rho_c(z)=3H^2(z)/(8\pi G)$.}, $\Delta_C = 2500,
500, 200$. As detailed in the appendix, local density maxima around
which growing spheres encompassing a given overdensity, are searched
by assigning density at the positions of particles using 64 SPH neighbours and
without allowing halos to overlap with each other.

\subsection{Matching halos}

\label{sec:match}
Since all three simulations share the same initial
conditions, dark matter particles have the same progressive
identification number (IDs). We exploit this information to match
halos identified in different simulations. Using a given halo
identified in the
DM simulation as the reference, a halo in the CSF or AGN simulation is
defined to be the counterpart of the DM halo whenever it includes the
largest number of DM particles belonging to the latter.  We define the
matching rate as the ratio of matched to total number of dark matter
particles in the DM halo. Clearly, the larger this rate, the more
accurate is the matching. In order to avoid multi-matching, i.e. two
halos from CSF/AGN simulation matched to one halo in the DM one, only
halos with matching rate larger than $0.5$ are selected. We 
verified that the fractions of matched SO halos for $\Delta_c = 500$,
are $97.5\%$ at $z = 0$, $98.3\%$ at $z = 0.6$, $98.6\%$ at $z =
1.0$ and $99.4\%$ at $z = 2.2$. Most of the mismatched halos have
smaller halo mass, e.g. $85\%$ of them have halo mass $M_{500} <
10^{13} \hMsun$ at $z = 0$. 

At each overdensity $\Delta_c$, we only consider halos with
$M_{\Delta_c} \geq 10^{12.5} \hMsun$. With this choice, the smallest
halo can still have $\sim 1000$ particles within the corresponding
$R_{\Delta_c}$. However, to allow for a complete matching, we consider
halos as small as $M_{\Delta_c}=10^{11.5} \hMsun$ in the AGN and CSF
simulations to be matched to the halos in the DM simulation. As shown
by \cite{Reed2013}, halos resolved with fewer than $N \sim 1000$
particles are unlikely to be used for a high-accuracy HMF measurement.
Furthermore, \cite{Watson2013} also pointed out that the correction
for low number of particles sampling FoF halos from \cite{Warren2006}
is $\sim 2$ per cent for the FoF halos containing 1000 particles. 
We used a fixed mass bin $\Delta \log M = 0.2$ for the calculation of
the HMF, without further correction. As discussed by \cite{Lukic2007},
the uncertainty in the HMF resulting from the choice of the binning is
negligible as long as the bin width does not exceed $\Delta \log M =
0.5$.

\section{Results}
\label{results}

Basic information on the number of halos identified by the FoF and SO
finders can be obtained from the cumulative HMF. We just mention here
that over $10^4$ halos are always found with both methods at $z = 0$
with halo mass $M \geq 10^{12.5} \hMsun$.  This number can reach $\sim
70000$ for FoF halos and for SO halos with $\Delta_c = 200$. At the
highest considered redshift, $z = 2.2$, this number is still $\sim
10^4$ for FoF and for SO halos with $M_{200} \geq 10^{12.5}$. However,
at the same redshift we only have $\sim 10^3$ SO halos with $M_{2500}
\geq 10^{12.5}\hMsun$. The CSF simulation has more both SO and FoF
halos than the DM one at all redshifts and halo masses, an increase
that is less apparent for $\Delta_c = 200$.  On the contrary, the AGN
simulation produces fewer halos of fixed mass than the DM one. Due to
limited simulation box size, only a few halos have mass $M \geq
10^{15} \hMsun$ for FoF and SO with $\Delta_c = 200$. Given the
limited dynamical range accessible to our simulations, we attempt in
the following to provide fitting expressions to the corrections to the
HMF induced by baryon effects, while we avoid providing absolute
fitting functions to the HMF.

\subsection{The HMF from Friend-of-Friend}
%%%%%%%%%% FIG 1 %%%%%%%%
\begin{figure}
%\hspace{1.3 cm}
%\vspace{1.3 cm}
\includegraphics[width=0.5\textwidth]{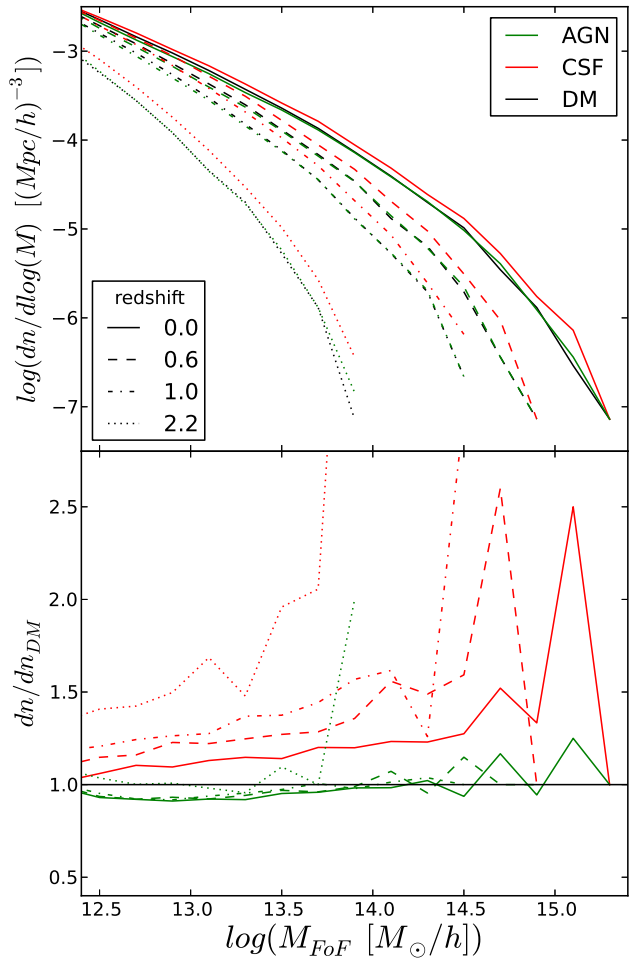}
\caption{Baryon effects on the halo mass function (HMF) for halos
  identified with the Friend-of-Friends (FoF) algorithm. Upper panel:
  FoF halo mass functions (HMFs). Different line styles are for
  different redshifts, as indicated in the legend in the bottom left
  corner, while different colours refer to the different simulations
  (legend in upper right corner). Lower panel: ratios between each of
  the HMFs from the hydrodynamical simulations and the HMF of the DM
  simulation.}
\label{fig:hmf_fof}
\end{figure}
%%%%%%%%%%%%%%%%%%%%%%%%%

We compare in the upper panel of Fig. \ref{fig:hmf_fof} the HMFs for
 the three different simulations, while the lower panel shows the
relative difference between each of the two hydrodynamical simulations
and the DM one.  As for the effect of the baryon physics described by
the CSF model, we note that the difference with respect to the DM case
has a clear redshift evolution and halo mass dependence. As redshift
decreases from $z = 2.2$ to 0, the HMF ratio drops from $\sim 1.6$ to
$\sim 1.1$, with a weak increasing trend of this ratio with halo mass,
at all redshifts. Quite remarkably, including AGN feedback has the
effect of reducing the difference with respect to the DM-only case:
the HMF ratio drops to about unity for massive halos with $M_{FoF}
\simgt 10^{14} \hMsun$, while at smaller halo mass it decreases to $\sim
0.9$ for $M_{FoF} \approx 10^{13} \hMsun$.  Unlike the CSF case,
these differences do not show any evidence of redshift evolution
from $z = 1$ to $z = 0.0$. At higher redshift, $z = 2.2$, the HMF
ratio keeps fluctuating around 1, as a consequence of the limited
statistics of halos due to the finite box size.

\subsection{The HMF from Spherical Overdensity}
%%%%%%%%%% FIG 2 %%%%%%%%
\begin{figure*}
%\hspace{-8.5 cm}
%\vspace{0.2 cm}
\includegraphics[width=1.0\textwidth]{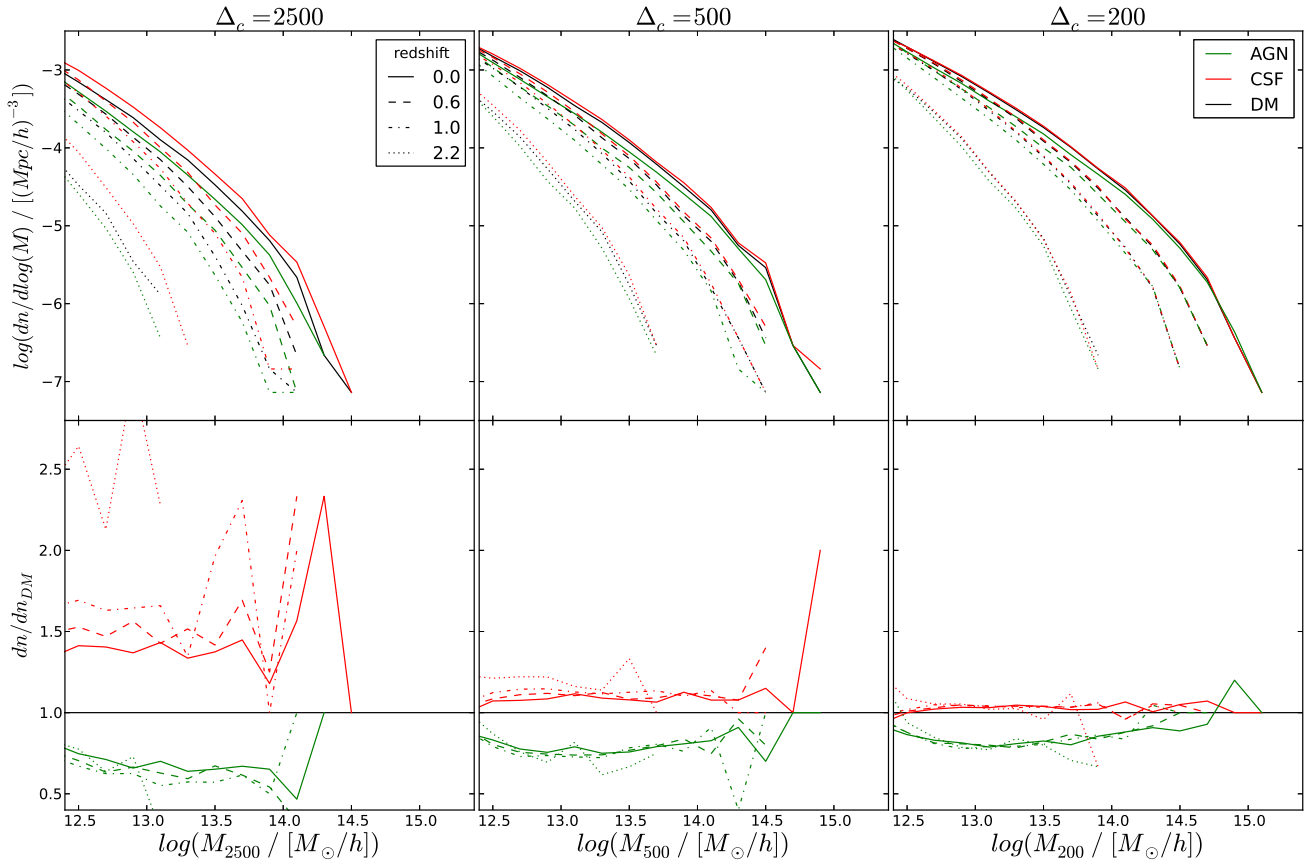}
\caption{Effect of baryons on the halo mass function (HMF) for
  spherical overdensity (SO) halos. Each panel is the same as in
  Fig. \protect\ref{fig:hmf_fof}, but for the spherical SO HMFs
  with halo masses computed for $\Delta_c = 2500$ (left panel), 500
  (middle panel) and 200 (right panel). }
\label{fig:hmf_so}
\end{figure*}
%%%%%%%%%%%%%%%%%%%%%%%%%

We compare in the upper panels of Fig. \ref{fig:hmf_so} the HMFs
obtained from the SO halo finder at three overdensities, $\Delta_c =
2500, 500, 200$ (from left to right), along with the ratios of the
HMFs from the CSF and AGN simulations with respect to the DM-only
result (lower panels).  As expected, baryons have a larger impact at
the highest considered overdensity, $\Delta_c = 2500$. In this case,
the ratio between CSF and DM HMFs shows a redshift evolution similar
to the FoF results but with a higher amplitude, ranging from $\sim
1.4$ at $z = 0$ to $\sim 2.5$ at $z = 2.2$, but with no significant
dependence on the halo mass. At lower overdensities, $\Delta_c = 200$
and 500, the redshift evolution becomes weaker and the differences
with respect to the DM case are reduced, with only a $\simlt 5$ per
cent difference at $\Delta_c=200$ \citepalias[similar results for the
CSF case were also found by][]{Cui2012a}.

When AGN feedback is included in the simulation, the corresponding HMF
drops below the HMF from the DM simulation, by an amount that
decreases for lower $\Delta_c$ values, with no evidence for redshift
dependence of the HMF difference. At $\Delta_c = 2500$, this ratio has
a weak halo mass dependence, ranging from $\sim 0.7$ at $10^{12.5}
\hMsun$ to $\sim 0.5$ at $10^{14} \hMsun$. At $\Delta_c = 500$ and
200, the difference between AGN and DM HMFs reduces, with a mild
dependence on halo mass: $dn/dn_{DM} \approx 0.7, \approx 0.8$ at $M
\approx 10^{13} \hMsun$ to $dn/dn_{DM} \approx 0.9, \approx 1.0$
($\Delta_c = 500, 200$, respectively) in the high mass end.

In general, the effect of including baryons on the HMF goes in the
same direction, independent of whether FoF or SO halo finders are
used. While this holds at a qualitative level, as expected
quantitative differences between FoF and SO results are found,
especially for the AGN case. As we will discuss in the following, the
effect of including AGN feedback is that of producing halos that are
less concentrated than in the CSF case. As a result, one expects that
matching SO and FoF HMFs requires in the CSF simulation a higher
$\Delta_c$ than in the AGN simulation. Many efforts are made to
rematch the two halo mass functions by tuning $b$ and $\Delta_c$
\citep[for example][]{Lukic2009, Courtin2011, More2011}. However, as
shown in \cite{Watson2013}, even in dark-matter-only simulations,
matching FoF HMFs to SO HMFs not only depends on the choice of $b$ and
$\Delta_c$, but also on the concentration parameter, pseudo mass
evolution, and the problems inside the two algorithms. These
quantitative differences between FoF and SO results make this matching
progress even more complex if baryon models are taken in account.

%%%%%%%%%% FIG 3 %%%%%%%%
\begin{figure*}
\includegraphics[width=1.0\textwidth]{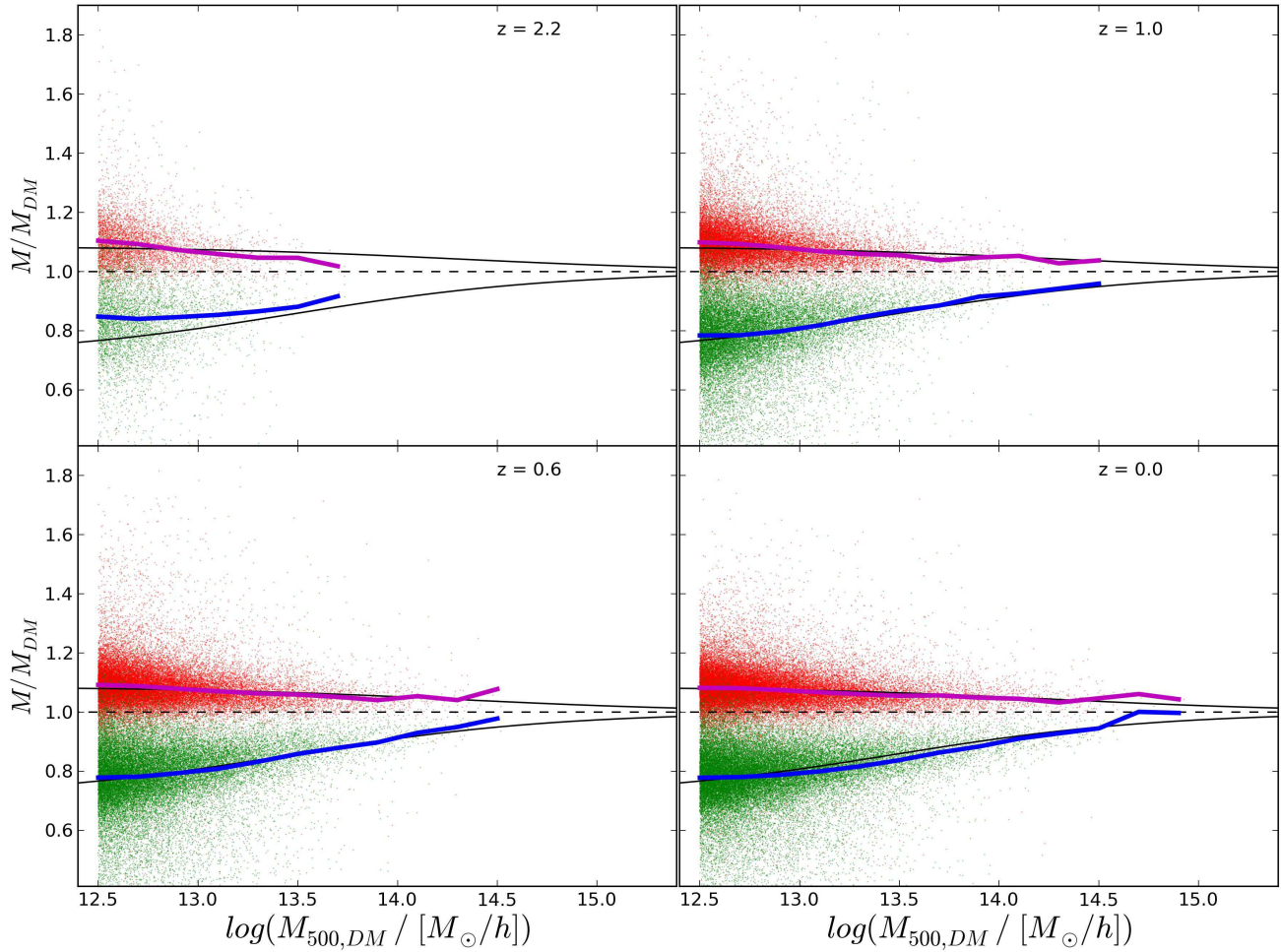}
\caption{Mass dependence of the ratio of masses of matched SO halos
  computed for $\Delta_c = 500$ at different redshifts, as reported in
  the upper--right corner of each panel. Each point represents a halo
  mass ratio between the matched CSF (red points) or AGN (green
  points) halos to DM ones, as a function of the mass of the matched
  halo in the DM simulation. The thick magenta and green lines show
  the mean values of this ratio within each mass bin for the CSF and
  AGN simulations, respectively. The best--fitting relation for the
  mass correction of Eq. \protect\ref{eq:1} is shown with the solid
  black lines.  We note that the same relation provides a good fit at
  all redshifts, at least up to $z=1$. See Table 1 for the values of
  the parameters defining this best-fit relation.  }
\label{fig:md_so}
\end{figure*}
%%%%%%%%%%%%%%%%%%%%%%%%%

In order to understand the origin of the baryonic effects on the HMFs
predicted by our simulations, we further focus on the difference of
masses of matched halos at overdensity $\Delta_c = 500$ (see Section
\ref{sec:match} for the description of the matching procedure).  We
show in Fig. \ref{fig:md_so} the ratio between masses of matched halos
in each one of the two hydrodynamical simulations and in the DM
simulation (red and green points for the CSF and AGN case,
respectively). In each panel, the thick lines show the mean value of
these ratios computed within each mass bin (magenta for CSF and blue
for AGN).  As for the CSF case, the effect of baryons is that of
increasing halo masses by an amount which is almost independent of
redshift. At each redshift, the halo mass ratio weakly decreases with
halo mass, from $\sim 1.1$ at $M_{500} = 10^{12.5} \hMsun$ to $\sim
1.05$ at $M_{500} \simgt 10^{13.5} \hMsun$, then becoming constant
\citepalias[see also][]{Cui2012a}.  As for the AGN simulation, the
effect of baryons goes in the opposite direction of decreasing halo
masses, thereby decreasing the corresponding HMF, as shown in
Fig. \ref{fig:hmf_so}. Also in this case, there is no evidence for a
redshift evolution of the halo mass ratio, at least below $z = 1.0$.
However, there is an obvious increase of this ratio with halo mass,
that ranges from $\sim 0.8$ at $M_{500} = 10^{12.5} \hMsun$ to $\simeq
1$ for the most massive halos found in our simulation box. Similar
trends are also found for the mass ratio with $\Delta_c = 2500, 200$,
both of which also show no evidence of redshift dependence for both
hydrodynamical simulations. We verified that using the median value of
those data points gives almost identical lines to these mean lines.
As discussed in the Appendix C, this effect of reduction of halo
masses in the presence of AGN feedback is quite robust against
numerical resolution. We refer to this Appendix for a more detailed
discussion of the resolution test that we carried out.

%%%%%%%%%% FIG 4 %%%%%%%
\begin{figure}
\includegraphics[width=0.5\textwidth]{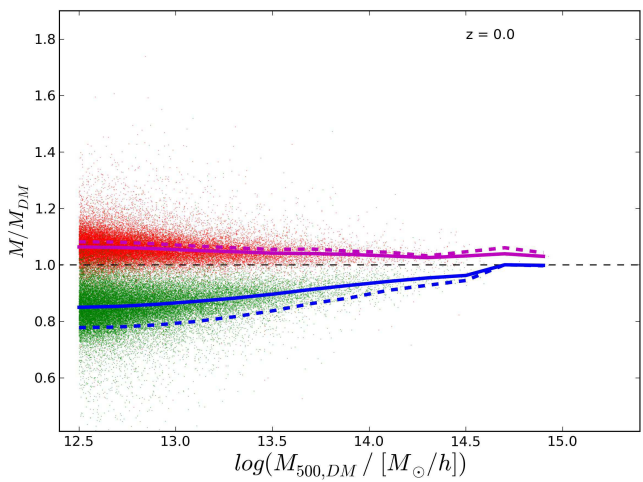}
\caption{The same as the bottom right panel of Fig. \ref{fig:md_so},
  but for halo masses in the CSF/AGN simulations computed within the
  $R_{500}$ radius of the corresponding halos from the DM
  simulation. The dashed thick lines are the previous results in
  Fig. \ref{fig:md_so} at $z = 0$.}
\label{fig:sr}
\end{figure}
%%%%%%%%%%%%%%%%%%%%%%%%%

Since the masses of SO halos are computed by adding up all the
particles within a sphere with radius $R_{\Delta_c}$, it is clear that
a change of the halo density profiles induced by the presence of
baryons would also change the corresponding values of
$R_{\Delta_c}$. In order to quantify the effect of this variation, we
also compute masses for each halo in the CSF/AGN simulations by using
the value of $R_{\Delta_c}$ of the corresponding halo identified in
the DM simulation.  In Fig. \ref{fig:sr}, we show again the halo mass
difference at $\Delta_c = 500$ after applying this re-tuning of the
halo radii. A comparison with the $z = 0$ result from in
Fig. \ref{fig:md_so} demonstrates that these ratios are only slightly
shifted towards unity, for both CSF and AGN models. This small change
implies that the differences in halo masses are mostly contributed by
the baryon effects on the halo density profiles, which can not be
recovered by simply changing the halo radius.

Including only SN feedback in the form of galactic ejecta is already
known not to be able to regulate overcooling at the centre of
relatively massive halos, with $M> 10^{12.5} \hMsun$. Adiabatic
contraction \citep[e.g.][]{gnedin2004}, associated to the condensation
of an exceedingly large amount of cooled gas, leads then to an
increase of density within a fixed halo aperture radius and,
therefore, to an increase of the halo mass with respect to the DM
case. The opposite effect is instead associated to the inclusion of
AGN feedback. In this case, the sudden displacement of large amount of
gas at epochs corresponding to the peak of AGN feedback efficiency,
taking place at $z\sim 2$--3, causes sudden variations of the halo
potential, which reacts with an expansion, thus decreasing halo masses
(see discussion in Sect. \label{gnedin2004} below).

\subsection{Density profiles}
\label{sec:profs}
Having quantified the variation of halo masses, we now discuss how
this variation is associated to changes in the total density profiles
of halos induced by baryonic processes.

%%%%%%%%%% FIG 5 %%%%%%%%
\begin{figure*}
\includegraphics[width=1.0\textwidth]{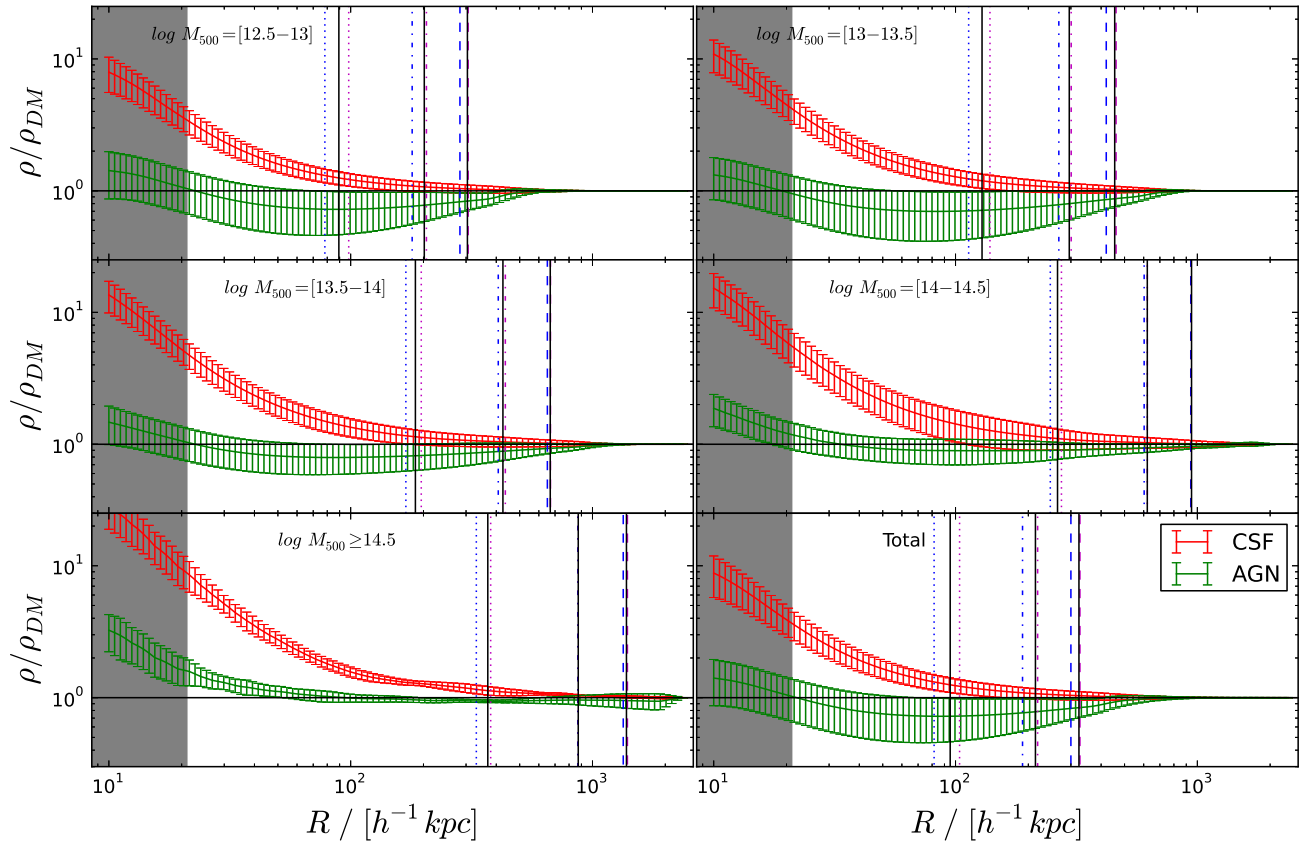}
%\hspace{4 in}
%\vspace{1.5 cm}
\caption{The stacked ratios of cumulative density profiles,
  $\rho(<R)$, of matched halos identified in each one of the
  hydrodynamical simulations and in the DM one. Different panels
  refer to different halo mass ranges, with the bottom right panel
  showing the results for all halos with mass $M_{500}\ge 10^{12.5}
  \hMsun$.  In each panel, the red and green solid lines are for the
  mean density ratios for the CSF and AGN case, respectively. Error
  bars indicate the $1 \sigma$ scatter around the mean.  Vertical
  dashed, dot--dashed and dotted lines indicate the median values of
  $R_{200}$, $R_{500}$ and $R_{2500}$ computed within each halo mass
  interval, with magenta and blue colours referring to the CSF and AGN
  simulations, respectively. We also show with continuous black lines
  the median values of $R_{2500}$, $R_{500}$ and $R_{200}$ for the DM
  simulation. The shadow regions show the limits of the gravitational
  softening.}
\label{fig:dpt}
\end{figure*}
%%%%%%%%%%%%%%%%%%%%%%%%%

To this purpose, we show in Fig. \ref{fig:dpt} the stacked ratio
between density profiles for halos belonging to different mass
intervals. We consider in this plot halos that are matched in the
CSF/AGN and in the DM simulations. Halos are separated into five mass
bins, according to the $M_{500}$ halo mass, measured in the DM
simulation. For each matched halo, we first compute the ratio of
cumulative density profiles, $\rho(<R)$. Then, we stack the density
profile ratios for all halos belonging to the same mass bin. The
stacked density profile ratios for CSF and AGN halos are shown with
red and green curves, respectively, with error bars indicating the $1
\sigma$ intrinsic scatter within each mass interval.

We note that density profiles for CSF halos are always higher than
those for the DM simulation for all mass bins. The effect is stronger
towards the cluster centre, as a result of adiabatic contraction
triggered by gas condensation from cooling. This result, which is in
line with those found by several previous analyses
\citep[e.g.][]{gnedin2004,puchwein2005}, indicates that the feedback
included in our CSF model is not efficient in counteracting the effect
of radiative cooling in increasing total density in central halo
regions. While the effect is quite small at $R_{200}$, it becomes
sizeable at $R_{2500}$, where density increases by up to about 50 per
cent for the smallest resolved halos.

It is quite interesting that a different behaviour is found when AGN
feedback in included. In this case, an increase of total density
associated to gas condensation is only found at small radii, below
(20--30) $h^{-1}$kpc, which are however close the smallest scale that
can be trusted at the resolution of our simulations. At larger radii
the effect of AGN feedback is that of decreasing the halo density
profile, an effect that becomes negligible at large radii, approaching
$R_{200}$, and for the most massive halos found in our
simulations. This result is in line with those from previous analyses
of cluster simulations including AGN feedback
\citep[e.g.][]{Duffy2010,Mead2010,Dubois2010,Killedar2012,Martizzi2013,Cui2014}. The
decrease of density profiles is caused by the effect of AGN feedback
that, at redshifts corresponding to the peak of BH accretion $z\simeq
2$--3, causes a sudden expulsion of gas from the potential wells of
the cluster progenitor halos \citep[similar result is also found by
][]{Dubois2010}. The expulsion of large amount of gas causes potential
wells to react with some expansion, thereby causing a decrease of the
density in central regions.

\subsection{Correcting the halo mass function}

%%%%%%%%%% FIG 6 %%%%%%%%
\begin{figure*}
\includegraphics[width=1.0\textwidth]{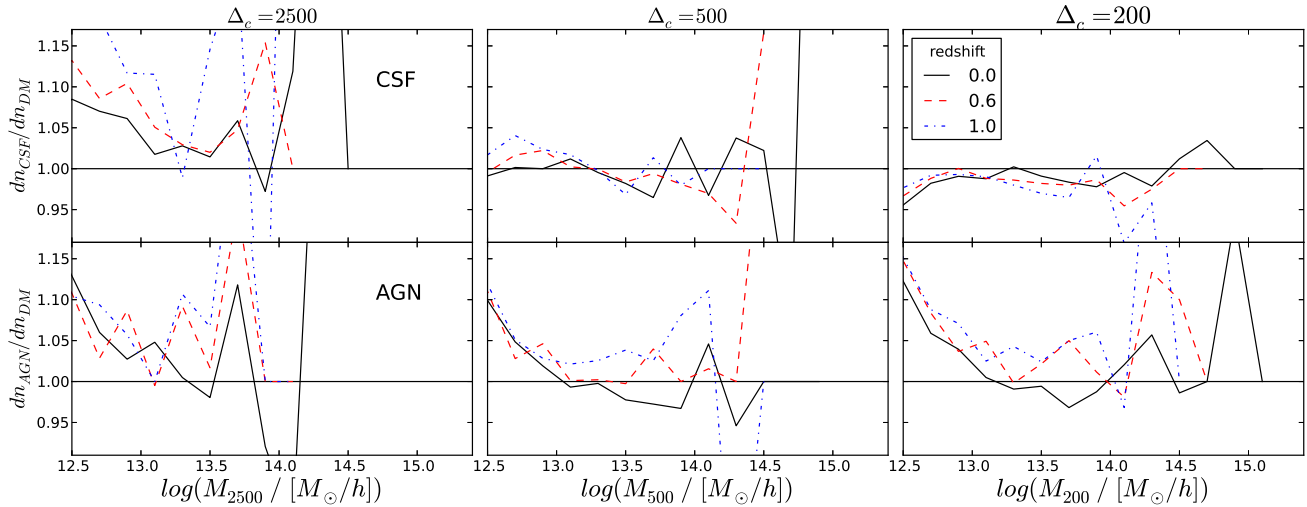} 
%\hspace{10cm}
%\vspace{2.5 cm}
\caption{The ratios between the halo mass functions from the CSF (upper
  panel), AGN (lower panel)
  simulations and from the DM simulation, after applying the correction
  to DM halo masses, as described in the text.  From left to right, we
  show results for overdensities $\Delta_c = 2500, 500, 200$. Different line
  styles and colors indicate different redshifts, a shown in the
  legend in the upper right panel.}
\label{fig:cd}
\end{figure*}
%%%%%%%%%%%%%%%%%%%%%%%%%

%%%%%%%%%%%%%%%%%%%%%%%%%%
\begin{table}
\begin{centering}
\begin{tabular}{l||c||c||c|}
\hline
 & $\Delta_c = 2500$ & $\Delta_c = 500$ & $\Delta_c = 200$ \\
\hline
\hline
  AGN & & & \\
\hline
$M_0$ & 13.952 & 13.471 & 13.334\\
\hline
$\alpha$ & 0.322 & 0.288 & 0.274\\
\hline
\hline
  CSF & & & \\
\hline
$M_0$ & 13.629 & 14.305 & 14.182\\
\hline
$\alpha$ & 0.295 & 0.085 & 0.045\\
\hline
\end{tabular}
\caption{Values of the best--fitting parameters describing the
  variation of mass of matched halos in the AGN, CSF and in the DM
  simulation, as described by Eq. \ref{eq:1}.}
\end{centering}
\label{Tb:fit}
\end{table}
%%%%%%%%%%%%%%%%%%%%%%%%%%%

Having traced the origin of the variation of halo masses induced by
baryonic processes, we investigate now whether the application of a
suitable correction to halo masses allows one to recover the HMF from
a hydrodynamical simulation, starting from the DM--only HMF. In
\citetalias[][]{Cui2012a}, we have already shown that the HMFs from
non--radiative and CSF simulations can be corrected to the
dark-matter-only HMF one, up to a residual scatter of $\simlt 3$ per
cent, by adopting a constant halo mass shift. From
Fig. \ref{fig:md_so}, we note that the mean values of the halo mass
difference between AGN and DM simulations (thick blue line) has a
significant mass dependence that needs to be included in the
correction.

As a convenient relation to describe the mass--dependence of the mean
mass ratio, we use these sigmoid functions
\be
\left\{ \begin{array}{l l}
f(x) \,=\, \alpha/(1+e^{-3 (x-M_0) / 2}) + 1 - \alpha & \quad \text{AGN},\\
f(x) \,=\, \alpha/(1+e^{3 (x-M_0) / 2}) + 1 & \quad \text{CSF},
\end{array} \right.
\label{eq:1}
\ee 
with $M_0$ and $\alpha$ considered as fitting parameters. Here, $f(x)$ is
the halo mass ratio between AGN and DM sets, with $x=\log
M_{\Delta_c,DM}$. Since the halo mass difference shows no evidence for
redshift evolution, at least for $z \leq 1.0$, we do not attempt to
fit a possible redshift dependence of the $M_0$ and $\alpha$ parameters and
exclude the results at $z=2.2$ from the analysis.
The values of $M_0$ and $\alpha$ obtained in this way are reported in Table
1 for the three values of the overdensity $\Delta_c$ at
which SO halos have been identified.
The thin black line shown in Fig. \ref{fig:md_so} indicates this
fitting to the correction for $\Delta_c=500$. We verified that a
redshift--independent mass correction also holds for the other
$\Delta_c$ values up to redshift $z = 1$. Further, we also checked
that the fitted parameters show no significant difference, whether we
choose to use median mass ratio or the mean mass ratio. 

We show in Fig. \ref{fig:cd} the ratio between the HMFs from the CSF,
AGN and from the DM simulations, after correcting halo masses in the
former according to Eq. \ref{eq:1}. In each panel, different
line-types indicate results at different redshifts with different
panels referring to different values of $\Delta_c$.  We note that the
correction to halo masses allows one to recover the DM HMF to good
accuracy, at least for masses larger than $10^{13} \hMsun$, the HMFs
are matched to the DM one with no apparent systematic bias within
random oscillations of $\simlt 5$ per cent. We note that results
become noisy whenever the sampling effects become important due to the
limited halo statistics. This is the case for large masses and,
especially, for the highest overdensity $\Delta_c=2500$. At small
masses, below $10^{13} \hMsun$, we note that the adopted mass
correction systematically produces an overestimate of the corrected
HMF for AGN set. For the smallest considered mass, $M_{DM} \ge
10^{12.5} \hMsun$, this overestimate is as large as 10--15 per cent,
the exact value depending on the overdensity $\Delta_c$. The reason
for this lies in the fact that this fitting of Eq. 1 does not provide
an accurate description of the mass correction at small halo masses
due to the halo mass cut at $10^{12.5} \hMsun$, as also shown in
Fig. \ref{fig:md_so}.

\section{Discussion and Conclusions}
\label{concl}

Based on a set of large--scale N-body and hydrodynamical simulations,
we presented an analysis of the baryon effects on the halo mass
function, with emphasis on the role played by gas accretion onto
super-massive black holes (SMBHs) and the ensuing AGN feedback. As
such, this analysis extends our previous one, presented in
\citetalias{Cui2012a}, to the case in which one accounts for a
suitable model for AGN feedback that regulates star formation within
massive halos, thereby improving the degree of realism of the
simulated galaxy clusters and groups. We compared three simulations: a
first one including only collisionless Dark Matter; a second one
including hydrodynamics with radiative physics and supernova (SN)
feedback; a third one which also includes AGN feedback. Based on these
simulations, we analyse how the halo mass distribution reacts to
overcooling and, in the presence of AGN feedback, to the sudden
displacements and expulsion of large amount of gas.

We use both Friends-of-Friends (FoF) and Spherical Overdensity (SO)
algorithms to identify halos. FoF halos are identified using a
standard (on-the-fly) FoF finder with a linking length $b = 0.16$. As
for SO halos, they are identified using an efficient memory-controlled
python code --- {\small PIAO}, in which halos are not allowed to
overlap. We focus on three overdensities $\Delta_c = 2500, 500, 200$
for the SO halos.

The main results of our analysis are summarised as follows.
\begin{description}

\item[1.] Including AGN feedback in hydrodynamical simulations causes
  a reduction of the halo mass function (HMF) with respect to the
  DM--only case, by an amount which depends on the overdensity within
  which halo mass is measured. This effect amounts to about 20 per
  cent for overdensity $\Delta_c=500$, with no evidence of mass and
  redshift dependence, and increases at higher overdensity. Therefore,
  our model of AGN feedback reverses the effect of radiative physics with
  no efficient feedback, which instead leads to an increase of the HMF.
\item[2.] The baryonic effects on the HMF can be traced to the effect
  that different feedback models have on the halo density profiles and
  total masses. In the absence of AGN feedback, we confirm that halo
  density profiles steepen as a consequence of adiabatic contraction
  associated to overcooling, thus causing an increase of halo
  masses. AGN feedback generates shallower density profiles and a
  corresponding decrease of halo masses. This effect is caused by the
  improved regulation of overcooling associated to AGN feedback and to
  the sudden displacement of large amount of gas, which makes the
  gravitational potential responding with an expansion. The relative
  decrease of halo masses is larger in smaller objects, where AGN
  feedback is relatively more efficient, and at higher overdensities;
  for $\Delta_c=500$, it amount to $\sim 20$ per cent at $\log
  M_{500}=12.5$, while becoming negligible for the largest halos found
  in our simulation box, with $\log M_{500}\simeq 15$. Interestingly,
  this effect is independent of redshift, at least up to $z=1$ where
  we have a large enough statistics of massive halos.
\item[3.] We provide a mass--dependent fit to the halo mass variations
  induced by baryonic effects. After applying this model for the mass
  correction to the HMF obtained from the DM--only simulation, we
  recover the HMF from hydrodynamical simulations, up to a random
  scatter of $\simlt 5$ per cent for halos more massive than $10^{13}
  \hMsun$, with no significant bias and independently of redshift.
\end{description}

Our analysis demonstrates that baryon effects can cause sizeable
variations of the HMF and, therefore, affects the measurement of
cosmological parameters from redshift number counts of galaxy
clusters. For instance, recent results from the Planck Collaboration
\citep[cf. also \citealt{Spergel2013}]{PlanckXXI} highlights that the
cosmological model preferred by CMB analysis over-predicts the number
of clusters expected in the SZ Planck Cluster Survey by about 50 per
cent. Interestingly, this tension would be relaxed, although by a
small amount, if the HMF calibrated with our AGN simulation is used to
predict SZ cluster number counts.

In a recent paper, \cite{Khandai} also investigated the effect of
including AGN feedback on the HMF. Since they considered simulation
boxes smaller than our, with size of 100$\hMsun$, they better probed
the low-mass end of the HMF, while having a worse sampling of the high
mass end. As a result of their analysis, they found that the FoF halo
mass function can be described with a universal form to a reasonable
accuracy, even after accounting for baryon effects.\footnote{After the
  submission of our paper, a paper by \cite{Velliscig14} appeared on
  the arXiv, which also included an analysis of baryon effects on halo
  masses and HMF, when AGN feedback is also included. Using box size
  and resolution quite similar to those of our simulations, they
  confirmed our results on the effect of AGN feedback on the HMF.}

Numerical convergence in the calibration of the HMF from purely
collisionless simulations could in principle be reached by ``brute
force'' (i.e. increasing the dynamic range accessible and the
gravitational force integration accuracy). The same may not be true
when the effects of baryons are included. In fact, our analysis shows
that baryons can generate variations of the HMF with respect to the DM
case which goes in different directions, depending on the nature and,
possibly, on the numerical implementation of feedback. In this
respect, confidence in the calibration of baryon effects in the HMF is
strictly related to the level of agreement between observational
results and model predictions for a variety of properties of galaxy
clusters. Current implementations of AGN feedback in cosmological
simulations produce cluster populations with an increased degree of
realism \citep[e.g.][ and references therein]{Puchwein2008,Dubois2012,
  Planelles2013, Planelles2014, LeBrun2014}. As an example, we compare
in Appendix \ref{A:sbf} two basic properties involving baryons in
clusters, namely the stellar mass fraction and the total baryon mass
fraction, with observational results. The good agreement between the
AGN simulation and observational results in quite encouraging. Still,
it is fair to admit that important tensions still exist between a
number of observed and simulated cluster properties, such as the
thermal structure of cool cores and the properties of the Brightest
Cluster Galaxies \citep[e.g.][ cf.  \cite{Kravtsov2014}]{Dubois2011,
  Ragone2013, Planelles2014}. There is no doubt that providing an HMF,
that accounts for the inclusion of the baryon physics at the percent
level of accuracy required by the next generation of large--scale
cluster surveys, still requires substantial work and, ultimately, a
precise numerical description of galaxy formation in a cosmological
framework.

\section*{Acknowledgements}
We thank the anonymous referee for a careful reading of
the manuscript and for useful comments, and
 Francisco Navarro--Villaescusa for his help in using Python
and MPI4Py. All the figures in this paper are plotted using the python
matplotlib package \citep{Hunter:2007}. Simulations have been carried
out at the CINECA supercomputing Centre in Bologna, with CPU time
assigned through ISCRA proposals and through an agreement with the
University of Trieste.  WC acknowledges a fellowship from the European
Commission's Framework Programme 7, through the Marie Curie Initial
Training Network CosmoComp (PITN-GA-2009-238356), and the supports
from ARC DP130100117, from UWA Research Collaboration Awards and from
the Survey Simulation Pipeline (SSimPL;
{\texttt{http://www.ssimpl.org/}}). We acknowledge financial
support from the PRIN-MIUR 2009AMXM79 Grant, from a PRIN-INAF/2012
Grant, from the PD51 INFN Grant and from the ``Consorzio per la Fisica
di Trieste''.

%*****************************************************************************
\bibliographystyle{mnras}
\bibliography{bibliography}

\appendix

\section{PIAO}
\label{A:Piao}

In this Appendix, we present a simple strategy to overcome a possible
memory limitation problem that can occur when dealing with
high-resolution, large-volume simulations. We then incorporate this
strategy in a standard Spherical Overdensity (SO) halo finder
algorithm and describe our {\rm P}ython spher{\rm I}c{\rm A}l {\rm
  O}verdensity code --- {\small PIAO}\footnote{{\small PIAO} is
  publicly available at https://github.com/ilaudy/PIAO}. We check the
consistency of {\small PIAO} and tackle the problem of halo overlapping
in SO methods in subsection \ref{con}. Finally, we also show that
{\small PIAO} is very efficient in memory control.

\subsection{Methodology and program}
\label{methods} 

With the rapid growth of supercomputing power, cosmological
simulations increase not only in resolution, but also in sheer
volume. Thus, the output dimension of these simulations increases
enormously, up to several Terabytes, or even Petabytes. Analysing such
a huge amount of data on a relatively small server, easily meets a
memory shortage problem.  Limited memory forbids reading all the
simulation information at one time.
 
We use a simple strategy to overcome this problem: splitting the whole
simulation box into small mesh-boxes, then analyzing them
one-by-one. We apply this strategy in two steps. Step one: the whole
box is meshed into small ones. Each mesh-box is written into separated
files which only contain only the needed information. Step two: each
file is iteratively read in and analysed. Although this strategy
inevitable wastes time on IO processes and hard disk space, we show
below that this meshing/IO time is usually very short compared to the
time spent in actually performing the analysis.

We apply such a strategy to the SO method, and for this purpose we
wrote a python code --- {\small PIAO}. {\small PIAO} makes extensive
use of the NumPy library. We also made a modification of the cKDTree
package in the sciPy.spatial library, adding the SPH density
calculation in it. We adopt the same SPH kernel as the {\small GADGET}
code.  To achieve high performance, cKDTree implements the algorithm
described in \cite{Maneewongvatana1999} in Cython.  {\small PIAO} is
also parallel with a python MPI package (MPI4py) to speed up the
calculation by taking advantage of multi-core CPUs.  In its first
step, {\small PIAO} reads in particles' positions from simulation
snapshots part by part, and meshes them into small boxes according to
mesh size. Only particles' ID, position, and mass are saved into mesh
files. In a second step, {\small PIAO} reads in one mesh file at a
time, and builds a buffer region around this mesh-box by reading in
information from all nearby mesh-boxes. All the particles' densities
within the mesh-box and its buffer region are calculated using the
nearest neighbours $N_{nbs}$ of each particle and applying the
selected SPH kernel over them.

{\small PIAO} uses the following simple loop to identify SO halos:
\begin{itemize}
  \item[1] The particle with the highest SPH density is chosen as the center
    of a SO halo. 
  \item[2] The code iteratively finds a radius $R_{\Delta}$, centered
    on the above particle, and enclosing a fixed overdensity $\Delta
    = M(<R_{\Delta})/ (4 \pi / 3 R^3_{\Delta} ) / \rho_{crit}$.  If
    the center was within the current mesh-box, all properties of this halo
    are computed, and the halo is then saved.
  \item[3] All particles within the radius
    $R_{\Delta}$ are flagged and excluded from being new centers by themselves.
    If we do not allow halo overlapping, those particles are also
    excluded from belonging to further halos.
  \item[4] If the current halo contains less than a chosen number of
    particles $N_{cut}$, the cycle ends. All of the remaining
    particles will not belong to any halo.
  \item[5] Next particle with highest SPH density is selected, and the
    loop continues from step [2].
\end{itemize}

We parallelled {\small PIAO} using the MPI4Py package, to take
advantage of multiple-core architectures.  Since the analysis of each
mesh-box is independent, this part is completely parallel and does not
need any communication nor barrier.  One processor is used for
controlling and assigning tasks, i.e. mesh-cubes to free
processes. This makes {\small PIAO} very flexible and balanced. In
principle, it can run on any number of processors.

\subsection{Consistency Check}
\label{con}

In this subsection, we test {\small PIAO} on a simulation, having
$256^3$ DM ($M_{DM}=1.93 \times 10^7 \hMsun$) and $256^3$ gas
particles ($M_{gas}=3.86 \times 10^6 \hMsun$) in a periodic box of
comoving size $18 h^{-1} {\rm Mpc}$. This simulation includes
metal-dependent cooling, star formation, and SN thermal and kinetic
feedback \citep[see more details about the model for star formation
and energy feedback in][]{Muppi}. Also this simulation has been run
using the TreePM-SPH {\small GADGET-3} code with the same cosmological
parameters of the simulations presented in the main text of the
present work. We used this simulation instead of those described in
Section \ref{simulation} just because it contains a lower number of
particles, and the analysis is thus quicker. Results on properties of
galaxies and of diffuse baryons in that simulation will be presented
elsewhere. Here we mainly focus on our halo finder performance, and in
particular on two {\small PIAO} parameters: mesh-box size and SPH
neighbour. We fix the overdensity parameter to $\Delta = 500$, and
minimum number of particles per halo to $N_{cut} = 64$.

\subsubsection{Mesh-box size}
\label{mbs}

Since {\small PIAO} employs a mesh to split the whole simulation into
small boxes, we first need to check that such a splitting does not
influence the halo identification procedure.  We used two mesh-box
sizes ($3, 6 \Mpc$) and checked our results against those obtained
when the whole test simulation is analyzed without any splitting.

Given that the mesh-box size should not affect the results, we expect to
find the same halo masses for the three different mesh-box
sizes. Since we use the same SPH neighbours $N_{nbs} = 64$ for those
tests, particles at the centers of identical halos are expected to
have the same SPH density.

We decided that halos found in different analyses are the same one if
their center particles have the same particle ID. All halos above mass
$M_{500} = 1.5 \times 10^9 \hMsun$ are well matched between different
mesh-box sizes. We confirm that all the matched halos have the same
mass. Thus, mesh size does not change the properties of all identified
halos. We found the same result both allowing and not allowing halos
to overlap.

\subsubsection{SPH Neighbours}
\label{Nei}

%%%%%%%%%% FIG A1 %%%%%%%%
\begin{figure}
\includegraphics[width=0.5\textwidth]{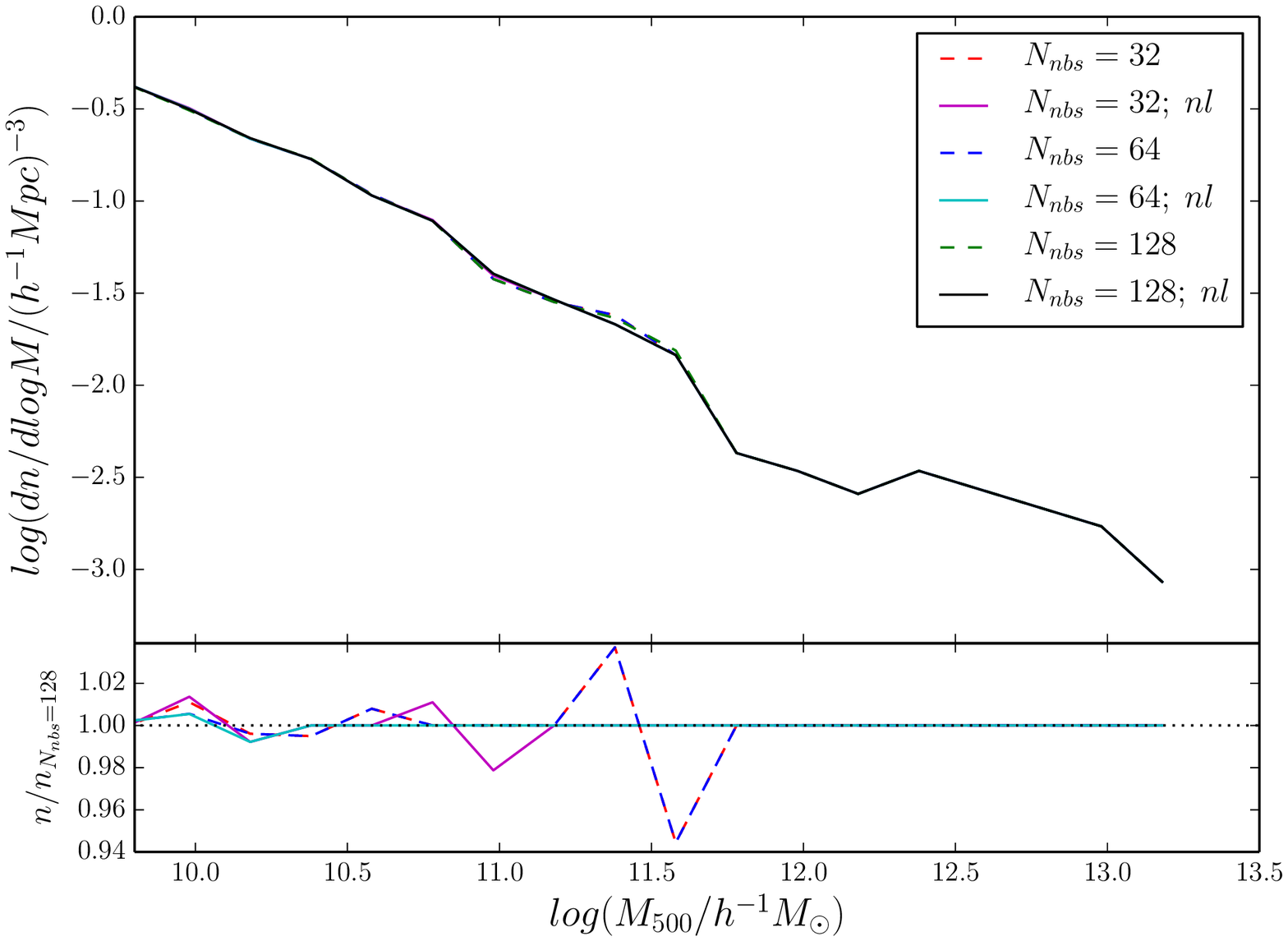}
%\vspace{3.5cm}
\caption{Halo mass functions at $\Delta = 500$ using different number
  of SPH neighbours (different line colors, see legend). Dashed lines
  represent halos identified allowing overlapping, while solid lines
  ones identified without any overlapping (``nl'' in the legend). In
  the lower panel, rations of the halo mass functions with respect
  with that obtained using $N_{nbs} =128$ neighbours is shown (dashed
  lines for overlapping halos and solid lines for non-overlapping
  halos, respectively.)}
\label{fig:c2}
\end{figure}
%%%%%%%%%%%%%%%%%%%%%%%%%

%%%%%%%%%% FIG A2 %%%%%%%%
\begin{figure}
\includegraphics[width=0.5\textwidth]{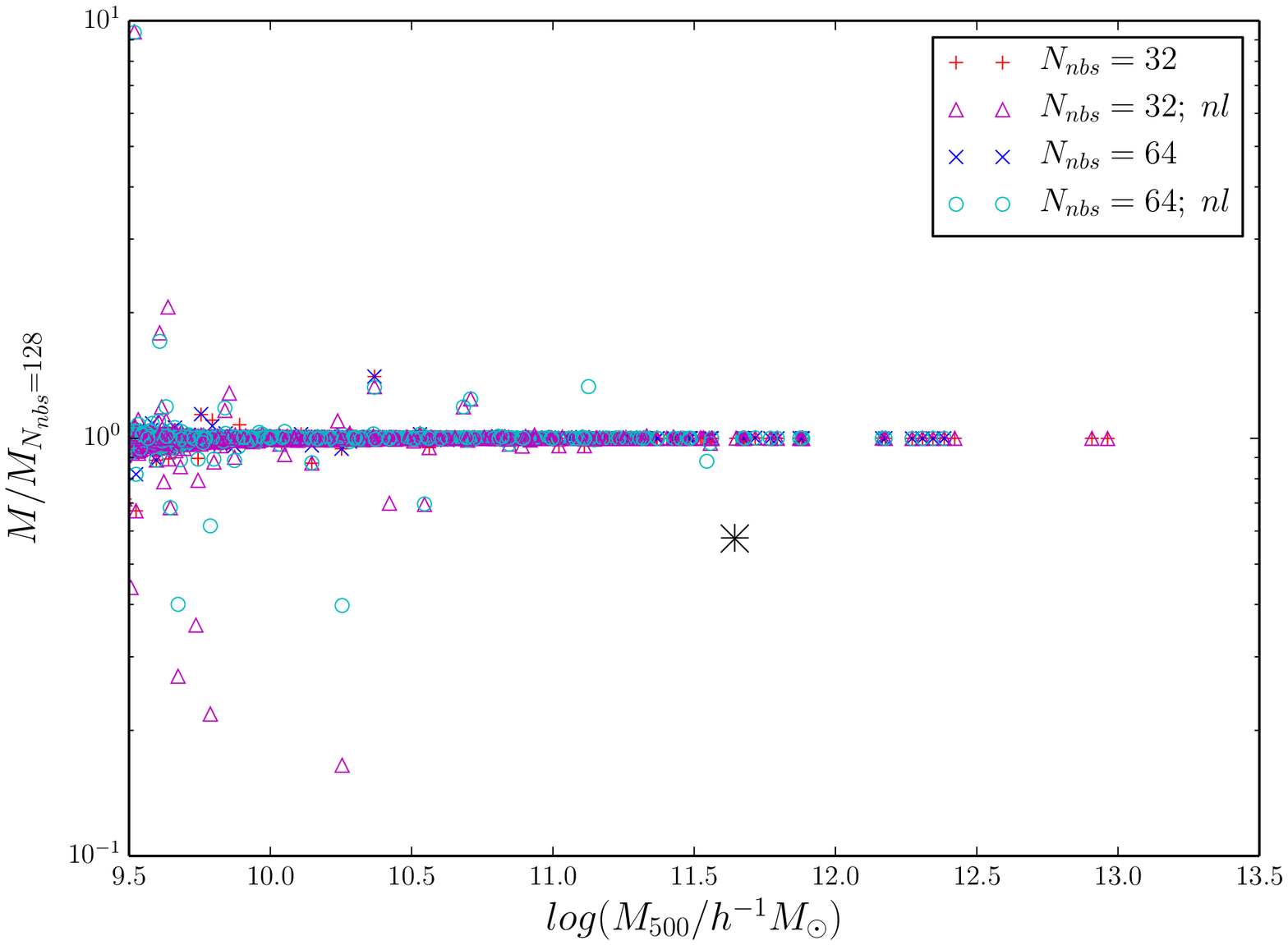}
%\vspace{3.5cm}
\caption{Ratio of halo mass function with respect to that obtained
  using halos identified with $N_{nbs} = 128$ SPH neighbours, as a
  function of halo mass $M_{500}$.  As shown in legend, different
  color symbols indicate halos identified with a different number of
  SPH neighbours. As in Figure \ref{fig:c2}, ``nl'' in legend
  indicates non-overlapping halos. Big black symbols indicate that the
  matched halos have a center offset larger than $50 ~ Kpc/h$.}
\label{fig:c2m}
\end{figure}
%%%%%%%%%%%%%%%%%%%%%%%%%

SPH density depends upon the number of neighbours $N_{nbs}$. Changing
the SPH density of particles can change the identification of the most
dense particle - i.e., the center of halo. If the center changes, all
halo properties can also vary.  We check the convergence of this
parameter by using three SPH neighbours numbers $N_{nbs} = 32, 64,
128$ on our test simulation, keeping fixed the mesh-box size to $6
\Mpc$.

In figure \ref{fig:c2}, we show halo mass functions from test
simulation with different SPH neighbours. The meaning of the color and
line-style is shown in the top-right legend. We show ratios of the
halo mass functions with respect to that obtained using $N_{nbs} =
128$ in the lower panel. At the high mass end of the mass functions,
$M_{500} \ga 10^{12} \hMsun$, there are no differences between the
results for any
value of $N_{nbs}$. This means that this SPH neighbour parameter has
no effect on massive halos. This SPH parameter makes the HMF ratio
fluctuate below a halo mass of $\sim 10^{11.5} \hMsun$. While at
smaller halo mass, these scatters are basically reduced within $\sim 1
\%$. Above all, we expect that the SPH density accuracy will have a
$\simlt 3 \%$ effect on halo mass function. If we don't allow halos to
overlap, the discrepancies are even smaller.

We also matched all the individual halos having mass $M_{500} > 1.5
\times 10^9 \hMsun$ and found using $N_{nbs} = 128$ neighbours, with
those identified using $N_{nbs} = 32$ and $64$ neighbours.  First, we
matched halos with the central particle having the same ID.  Then, for
all unmatched halos from our $N_{nbs} = 128$ analysis we calculated
the minimum distances from the centers of unmatched halos in the
$N_{nbs} = 32, 64$ analyses. If such a distance is smaller than both
the $N_{nbs} = 128$ and the $N_{nbs} = 32, 64$ halo radius, we decided
that the corresponding halos matches.  After this one-by-one matching
procedure, we are left with 7 ($0.15\%$) unmatched halos in our $N_{nbs} = 32$
analysis and 1 ($0.02\%$) with the $N_{nbs} = 64$ one when we allow overlapping.
In the case of non-overlapping halos, we have 5 ($0.11\%$) unmatched objects for
our $N_{nbs} = 32$ analysis and 2 ($ 0.04\%$) for the $N_{nbs} = 64$ one. Most of
the unmatched halos have $M_{500} < 10^{10} \hMsun$. We show the halo
mass ratios in figure \ref{fig:c2m} for all matched halos.

Even if the halo number, at a given mass scale, changes less than $3$
\% (as discussed above), inaccuracies in the SPH density evaluation
can lead to large halo mass difference for particular objects (see
magenta triangles in figure \ref{fig:c2m}). Therefore, when
halo-by-halo matching is needed, we recommend a higher number of
neighbours to be used in the density evaluation. 

\subsubsection{Overlapping Problem}
\label{op}

%%%%%%%%%% FIG A3 %%%%%%%%
\begin{figure}
\includegraphics[width=0.5\textwidth]{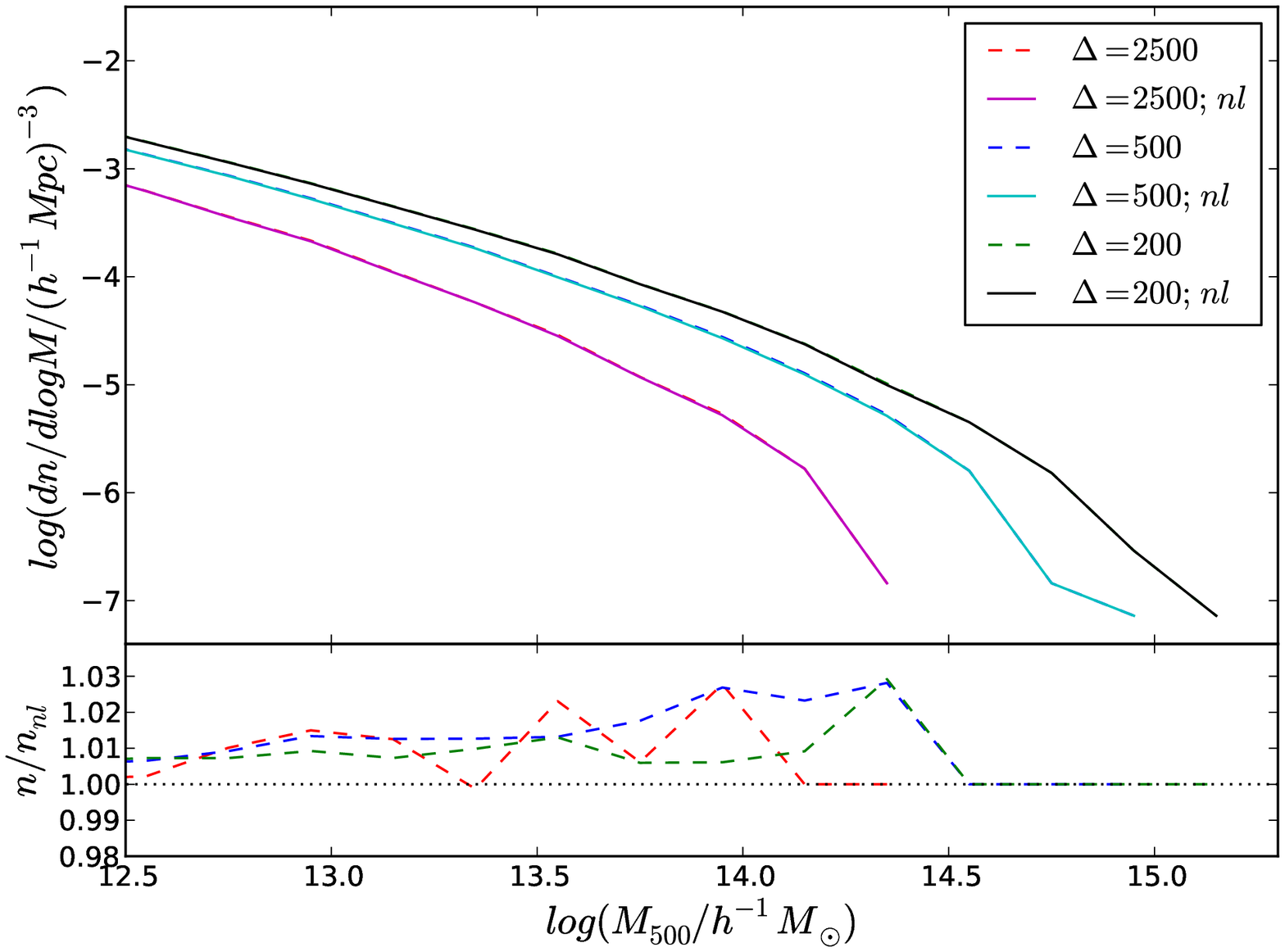}
%\vspace{1.7cm}
\caption{Halo mass functions at different
  overdensities $\Delta = 2500, 500, 200$. 
  The meanings of line colors and styles are shown in the top-right
  legend. Lower panel shows the ratio between overlapping an
  non-overlapping halo mass functions.
  When we allow halos to overlap the mass function is
  over-predicted by  $\simlt 3 \%$. This number has a
  weak dependence on halo mass. However, it seems
  consistent for the three overdensities. } 
\label{fig:c3dm}
\end{figure}
%%%%%%%%%%%%%%%%%%%%%%%%%

In X-ray or SZ observations, clusters are usually allowed to overlap,
and overlapping objects count as separate objects.  However,
usually such observations only focus on most massive clusters, with
masses larger than $10^{13} \Msun$. Overlapping between these
clusters is rare.  Even if two halos with such a mass overlap, at
worst one halo will have its mass increased of less than $50 \%$ with
the other one suffering a equivalent mass decrease.  Simulations
span a wider mass range, currently five or more orders of
magnitudes, and reach lower masses.  We need to have an accurate halo
mass function at per cent level over the whole covered mass
range. Thus, the difference between overlapping and non-overlapping
halo identification must be carefully examined.

To answer the question of how halo overlapping can change the halo
mass function, we used {\small PIAO} to analyze our DM-only simulation
described in section \ref{simulation} at
three overdensities $\Delta = 2500, 500, 200$, both allowing
overlapping and not allowing it. Since SPH neighbours should not
affect halo mass function too much (see the discussion in section
\ref{Nei}), we fixed the number of SPH neighbours to $N_{nbs} = 64$.
The minimum halo particle number parameter is set at $N_{cut} =
64$. Resulting halo mass functions are shown in figure
\ref{fig:c3dm}. In the lower panel of this figure, we show the ratios
of mass functions obtained at the various overdensities when we allow
overlapping with respect to the non-overlapping case. The effect on
the halo mass function is within 3 per cent over the whole halo mass
range, with the overlapping case systematically overpredicting the
mass function. This result does not strongly depend on the chosen
overdensity nor on the mass scale, apart from the higher masses ($M >
10^{14.5} \hMsun$, where the two analyses give identical
results.

\subsection{Summary}

We used a simple meshing strategy to overcome the problem that
analyzing large simulations on a small server or PC can be difficult
due to memory limitation, and present a simple parallel Python spherical
overdensity halo finding code --- {\small PIAO}.

{\small PIAO} employs two additional parameters besides the
overdensity $\Delta_c$: the mesh-box size, which splits the whole
simulation box into smaller ones, and the SPH neighbours number, that
is used for the SPH density calculation. In section \ref{con}, we
showed that the mesh-box size parameter does not influence the
identification of halos nor their properties. Since SPH density is
used to locate halo density peaks, we further investigated the impact
of the SPH neighbours number parameter on halo properties.  The halo
mass function is not strongly affected by it (at most at the $\simlt
3\%$ level).  On the other hand, one-by-one halo comparison suggests
that an inaccurate density estimate may lead to large disagreements
for individual halos. At last, we investigated the halo overlapping
problem in section \ref{op}, and showed that the halo mass function is
$\sim 1-3$ per cent higher for overlapping halos for three different
overdensities, independently from the value of the overdensity.

We notice here, that all these tests in this appendix are
done on a desktop PC, with a 4 core $2.67 GHz$ CPU and total memory
$\sim 3.4GB$. For test simulation with mesh-box size $6 \Mpc$ buffer
region, peak using memory is $\simlt 0.9GB$ for one 
processor. The meshing time is about 2 per cent of the analyzing
time. For the DM simulation, we used mesh-box size $41 \Mpc$ and 
$4.5 \Mpc$ buffer region. Although the DM
simulation has about 8 times more particles than the test simulation, the
allocated memory for one CPU is only $\sim 10 \%$ of the total
memory, because the DM simulation was separated into
1000 mesh boxes. However, using the same number of CPUs, the analyzing time for
the DM simulation is about 8 times more than for the test simulation.

\section{The stellar and baryon mass fraction}
\label{A:sbf}

%%%%%%%%%% FIG B1 %%%%%%%%
\begin{figure*}
\includegraphics[width=1.0\textwidth]{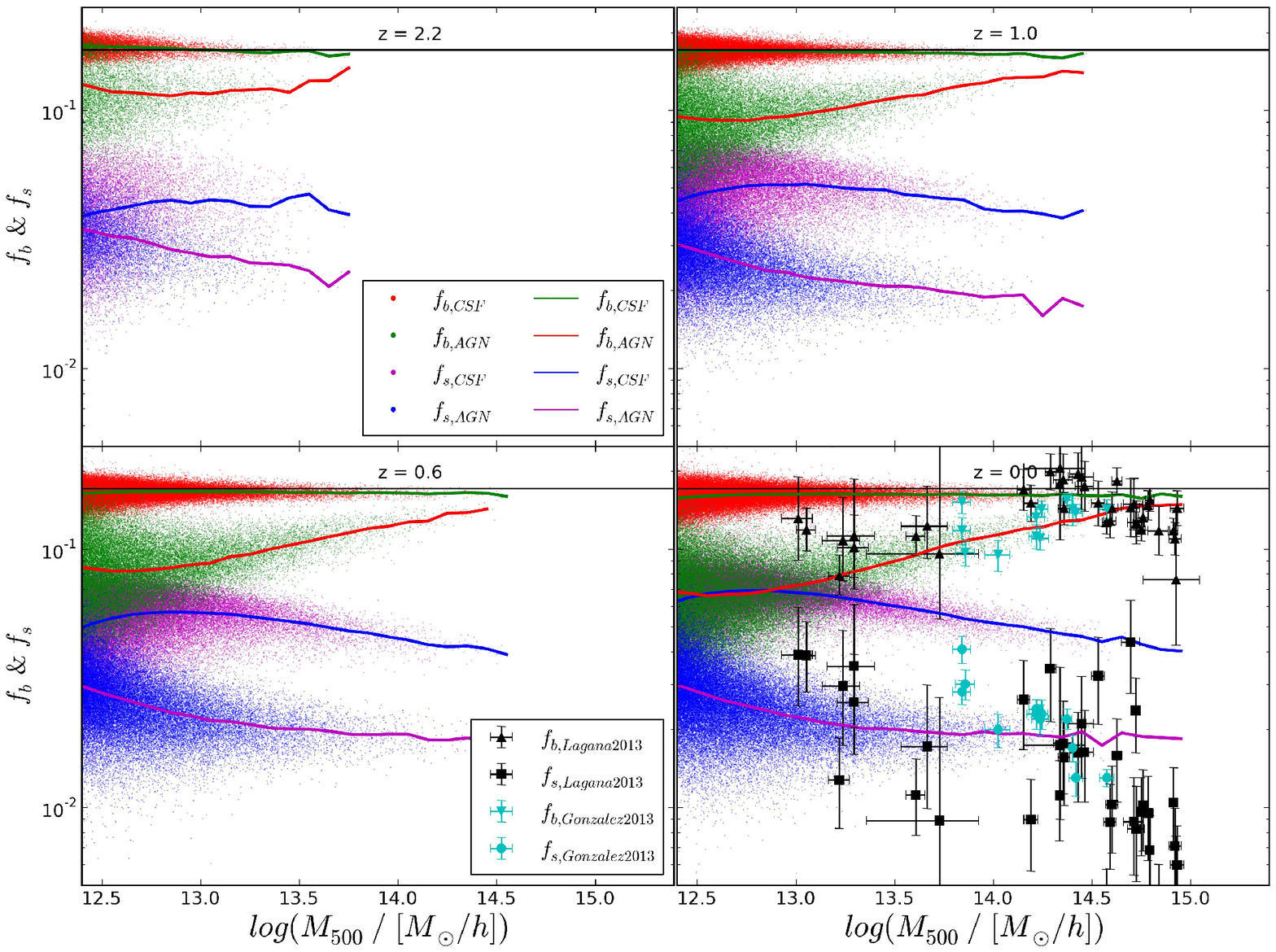}
\caption{Baryon and stellar mass fraction within $R_{500}$ for CSF and
  AGN simulations, described in Section \ref{simulation}. Different
  color points show baryon mass fraction ($f_b$) or stellar mass
  fraction ($f_s$) of each halo. Color lines are the mean value of the
  points. The line colors are reversed from points to highlight the
  difference, see the legend box in top-left panel for detail. The
  black horizontal line is the cosmic baryon mass fraction from our
  simulation. The three panels show the analysis at three different
  redshift $z=2.2$,$z=1.0$ and $z=0.6$.  In the lower-right panel, we
  compare the simulation results at redshift $z = 0.0$ to observations
  from \citep{Lagana2013, Gonzalez2013} (see the legend in lower left
  panel).}
\label{fig:fbfs}
\end{figure*}
%%%%%%%%%%%%%%%%%%%%%%%%%

For better understanding and calibrating the effects of baryons on the
HMF, the baryon and stellar fraction can provide a key element.  We
define the total baryon mass fraction as the ratio between the
gas+stars mass and the total mass within $R_{500}$: $f_b =
(M_{gas}+M_{star})/M_{tot}$, while the stellar fraction is $f_s =
M_{star}/M_{tot}$. The two fractions have been investigated in many
works \cite[from observation e.g.][]{Lin2003, Lin2012, Gonzalez2007,
  Gonzalez2013, Giodini2009, Andreon2010, Zhang2011, Lagana2011,
  Lagana2013}, \cite[or from theory e.g.][]{Ettori2006, Borgani2006,
  Fabjan2010, Puchwein2010, McCarthy2011,
  Planelles2013,LeBrun2014}. The total baryon fraction is commonly
found to increase with the mass of the halo, while the stellar
fraction seems to increase when going from clusters of galaxies mass
scale to groups ones.

In Fig. \ref{fig:fbfs}, we show the baryon fraction $f_b$ and stellar
fraction $f_s$ from CSF and AGN simulations, described in Section
\ref{simulation}. The four panels show results from redshift $z = 2.2$
to $z = 0$, Different color points show the fractions $f_b \& f_s$
computed for the two simulations, for each halo, while the
solid color lines are the mean of the points. Comparing the results
from CSF and AGN sets, we can see that without AGN feedback,
continuous star forming processes produce both higher stellar fraction $f_s$
and baryon fraction $f_b$ inside $R_{500}$ compared to the AGN set at
all redshifts. Similar to 
the finding of \cite{Planelles2013}, both fractions, for the two
simulations, show almost no redshift evolution for the most massive
halos. However, at smaller halo mass, $f_b$ computed on our CSF
simulation shows a smaller decrease in time, when compared to the
result from AGN simulation ($\sim 0.13$ at $z = 2.2$ to $\sim 0.07$ at
$z = 0$).  The stellar fraction $f_s$ computed on our CSF set
increases with time, while there is a slightly decreasing trend for
the same fraction calculated on our AGN simulation.

In the lower right panel of Fig. \ref{fig:fbfs}, we compare our
results with observations from recent works of \cite{Gonzalez2013,
  Lagana2013} at redshift $z=0$. Both papers use a WMAP 7
cosmology, and we corrected our results to account for that.  Clearly,
both fractions computed on our AGN simulation show a better match to
the observations than our CSF simulation results.  The trend of the
fraction $f_b$ with mass, in our AGN simulation, is also in good
agreement with observations. However, the fraction $f_s$ computed on
the same AGN simulation is flatter than the observation results at the
high mass end.  This indicates that AGN feedback in our simulation is
still not efficient enough to quench star formation at the observed
levels in the most massive halos.

\section{Resolution test}
\label{A:rt}

%%%%%%%%%% FIG C1 %%%%%%%%
\begin{figure}
\includegraphics[width=0.5\textwidth]{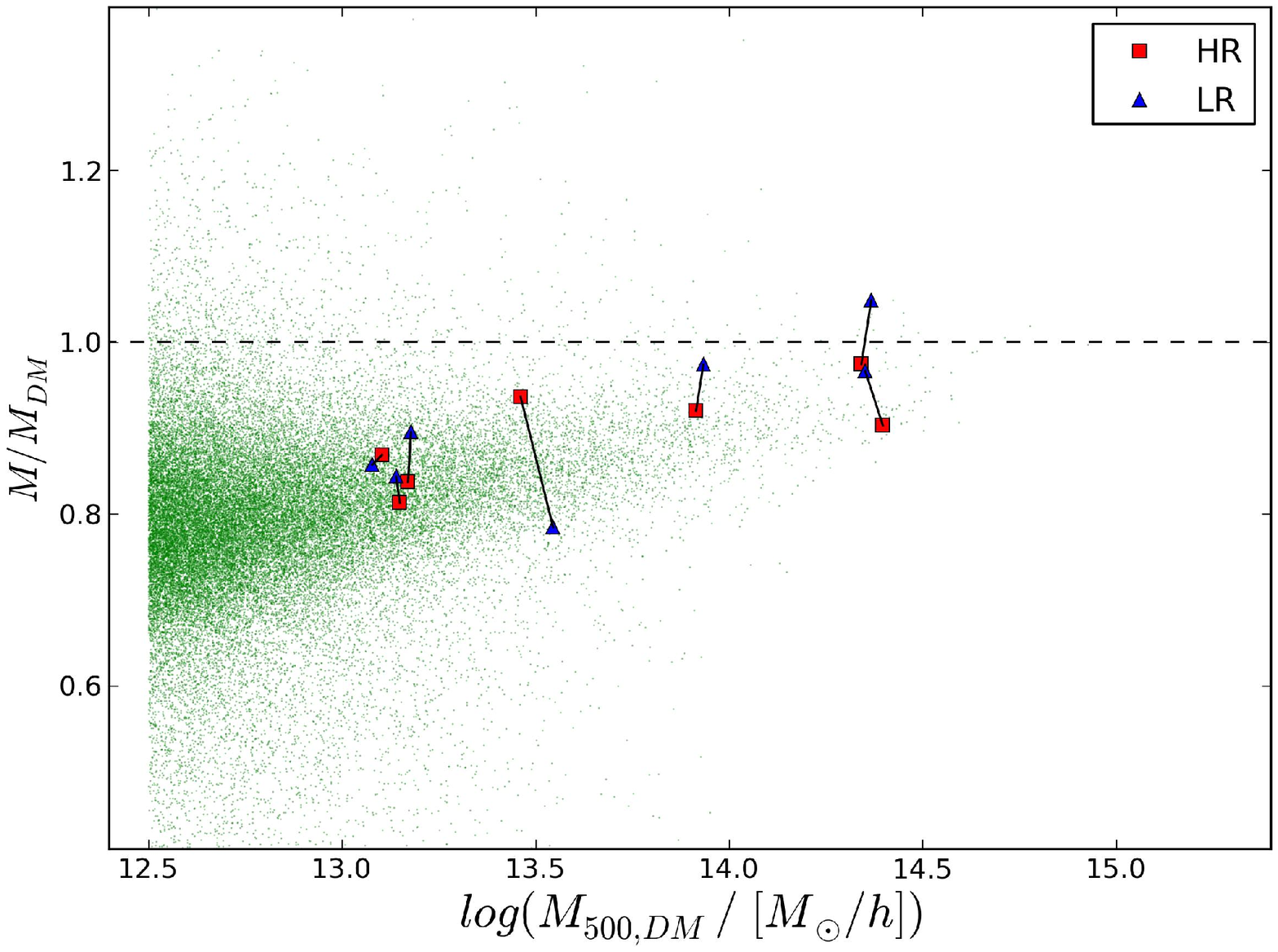}
\caption{Resolution test at $z=0$ for the values of the $M_{500}$ halo
  masses. Small green dots are for the halos identified in the
  simulations analyzed in this paper (see also bottom right panel of
  Fig. \ref{fig:md_so}. Overplotted are the results from zoomed-in simulation of
  galaxy clusters carried out for the DM and AGN cases, at two
  different resolutions. The lower resolution (LR; blue triangles)
  simulations have a mass resolution which is a factor of about 4
  better than for the reference cosmological boxes, while the higher
  resolution (HR; red squares) have a mass resolution 10 times better
  than the LR ones. Cluster counterparts identified in the two
  simulations sets are connected by black solid lines.}
\label{fig:rt}
\end{figure}
%%%%%%%%%%%%%%%%%%%%%%%%%

In this section we discuss the convergence against numerical
resolution of the results on the mass correction induced by the
baryonic effects included in the AGN simulations. While we did not
carry our simulations of cosmological boxes at higher resolution, we
analyzed zoomed--in simulations of galaxy clusters and groups carried
out at different resolutions. These simulations include all the
baryonic effects as the AGN large--box simulation analyzed in this
paper. More in detail, we used four of the 29 Lagrangian regions
surrounding massive clusters, presented by
\cite{Bonafede2011}. \cite{Ragone2013} and \cite{Planelles2014}, and 
presented results from hydrodynamical simulations of these Lagrangian
regions which include the effect as SN and AGN feedback, as in the
simulation considered in this paper. DM and baryonic particles in
these simulations have masses of $8.47\times 10^8\hMsun$ and
$1.53\times 10^8\hMsun$, respectively. As such they have
mass resolution which is a factor of about four better than the
cosmological simulations presented in this paper. Four of these
Lagrangian regions have been resimulated by further increasing mass resolution
by a factor of 10. We identified 7 halos in these four
simulations, all having masses of at least $10^{13}\hMsun$,
whose counterparts in the low--resolution (LR) and high--resolution
(HR) versions are both free of contaminant DM particles coming from
outside the zoomed-in Lagrangian regions. The $M_{500}$ values for these halos
have been compared to the corresponding masses measured in DM--only
version of the same simulations. The results on the mass variation
induced by the baryonic effects included in the AGN simulations are
shown in Figure \ref{fig:rt}, both for the LR and HR versions. 
Therefore, this figure illustrates how baryonic effects impact on halo
masses when mass resolution is increased by a factor of about 40.

Clearly, the relatively small number of objects prevents us from
drawing any robust statistical conclusion from this test. Still, we
confirm the decrease of halo masses when AGN feedback is
included. This decrease is also confirmed to be more pronounced in
smaller halos, with a trend that shows no evidence of resolution
dependence. This result is in line with the resolution test presented
by \cite{Velliscig14}.  More in detail, we note that the absolute
value of halo masses in the DM--only simulations (as reported on the
abscissa) can have small, but sizeable, variations with resolution,
again with no obvious trend. The reason for these variations lies in
the fact that, when mass resolution is increased, higher frequency
modes from the linear power spectrum are also added when computing the
displacement and velocity fields in the initial conditions. This
effectively causing small variations in the timing of halo formations
and mergers, which result in variations of halo masses at a fixed
redshift.

\bsp
\label{lastpage}
\end{document}